\documentclass[aps,preprintnumbers,floats,floatfix,superscriptaddress,preprintnumbers,showpacs,eqsecnum,nofootinbib,notitlepage,twocolumn]{revtex4-1}
\usepackage[utf8]{inputenc}
\usepackage{latexsym,array,theorem,mathrsfs,bm,float}
\usepackage{psfrag}
\usepackage{amsfonts,amsmath,amssymb,afterpage,epsfig,color,graphicx,tabularx,here,multirow,hyperref}

\newcommand{\nn}{\nonumber \\}

\newcommand{\Hdcal}{ {}^d \! \mathcal{H} }
\newcommand{\limitHcal}{ {}^3 \! \mathcal{H} }
\newcommand{\Hd}{ {}^d \! H }
\newcommand{\limitH}{ {}^3 \! H }
\newcommand{\Gdcal}{ {}^d \! \mathcal{G} }
\newcommand{\limitGcal}{ {}^3 \! \mathcal{G} }

\begin{document}
\baselineskip=12pt

\preprint{YITP-20-66, IPMU20-0050} 
\title{A consistent theory of $D\rightarrow 4$ Einstein-Gauss-Bonnet gravity}
\author{Katsuki Aoki}
\email{katsuki.aoki@yukawa.kyoto-u.ac.jp}
\affiliation{Center for Gravitational Physics, Yukawa Institute for Theoretical Physics, Kyoto University, 606-8502, Kyoto, Japan}

\author{Mohammad Ali Gorji}
\email{gorji@yukawa.kyoto-u.ac.jp}
\affiliation{Center for Gravitational Physics, Yukawa Institute for Theoretical Physics, Kyoto University, 606-8502, Kyoto, Japan}

\author{Shinji Mukohyama}
\email{shinji.mukohyama@yukawa.kyoto-u.ac.jp}
\affiliation{Center for Gravitational Physics, Yukawa Institute for Theoretical Physics, Kyoto University, 606-8502, Kyoto, Japan}
\affiliation{Kavli Institute for the Physics and Mathematics of the Universe (WPI), The University of Tokyo, 277-8583, Chiba, Japan}

\date{\today}

\begin{center}
\begin{abstract}
We investigate the $D\rightarrow 4$ limit of the $D$-dimensional Einstein-Gauss-Bonnet gravity, where the limit is taken with $\tilde{\alpha}=(D-4)\, \alpha$ kept fixed and $\alpha$ is the original Gauss-Bonnet coupling. Using the ADM decomposition in $D$ dimensions, we clarify that the limit is rather subtle and ambiguous (if not ill-defined) and depends on the way how to regularize the Hamiltonian or/and the equations of motion. To find a consistent theory in $4$ dimensions that is different from general relativity, the regularization needs to either break (a part of) the diffeomorphism invariance or lead to an extra degree of freedom, in agreement with the Lovelock theorem. We then propose a consistent theory of $D\rightarrow 4$ Einstein-Gauss-Bonnet gravity with two dynamical degrees of freedom by breaking the temporal diffeomorphism invariance and argue that, under a number of reasonable assumptions, the theory is unique up to a choice of a constraint that stems from a temporal gauge condition. 
\end{abstract}
\end{center}

\maketitle

\section{Introduction and summary}
Contrary to the common knowledge based on the Lovelock theorem~\cite{Lovelock:1970,Lovelock:1972vz}, a recent paper~\cite{Glavan:2019inb} intended to propose a novel $4$-dimensional covariant gravitational theory with only two dynamical degrees of freedom (dofs), by taking the $D\rightarrow 4$ limit of the Einstein-Gauss-Bonnet (EGB) gravity in $D>4$ dimensions. As is well-known, the Gauss-Bonnet (GB) term in $4$ dimensions is a total derivative and thus does not contribute to the equations of motion. An intriguing idea of \cite{Glavan:2019inb} is to multiply the GB term by the factor $1/(D-4)$ before taking the limit. It was shown that, at the level of equations of motion under a concrete ansatz of the metric, the divergent factor $1/(D-4)$ is canceled by the vanishing GB contributions yielding finite nontrivial effects. Despite the singular limit, it was conjectured that the $D\rightarrow 4$ limit should have only two dofs, based on the fact that the number of dofs of the $D$-dimensional EGB gravity is $D(D-3)/2$.

The original suggestion of the $D\rightarrow 4$ EGB gravity is in explicit contradiction with the common knowledge and hence came into questions. The papers~\cite{Lu:2020iav,Kobayashi:2020wqy} started with a direct product $D$-dimensional spacetime and then took the limit $D\rightarrow 4$. They found well-defined theories which belong to a class of Horndeski theory~\cite{Horndeski:1974wa} but with $2+1$ dofs, in general. The same results are deduced in~\cite{Fernandes:2020nbq,Hennigar:2020lsl} (see also~\cite{Mann:1992ar}) by adding a counter term in $D$-dimensions and then taking the $D\rightarrow 4$ limit. However, the scalar-tensor description lacks the quadratic kinetic term of the scalar field and thus suffers from the infinite strong coupling problem in general (see e.g.~\cite{Kobayashi:2020wqy,Bonifacio:2020vbk,Ma:2020ufk}). It was explicitly confirmed by \cite{Kobayashi:2020wqy} that the cosmological solution found by \cite{Glavan:2019inb} is infinitely strongly coupled in the scalar-tensor description. One has not to regard the cosmological solution as a solution of the scalar-tensor description since the cutoff of the theory is zero. The same pathology is expected to exist even around the black hole spacetime at least in its asymptotic region.
Another $D\rightarrow 4$ limit without the strong coupling was proposed by~\cite{Bonifacio:2020vbk} but with $2+1$ dofs again. The resultant theory is just the $(\partial\phi)^4$ theory. On the other hand, in~\cite{Gurses:2020ofy} it was shown at the level of equations of motion that a diffeomorphism invariant theory cannot be realized (see also~\cite{Ai:2020peo,Mahapatra:2020rds,Arrechea:2020evj}). They all show that there is no manifestly covariant novel $D\rightarrow 4$ EGB gravity with only two dofs, in agreement with the Lovelock theorem. Even if one adopts the scalar-tensor description, the spacetimes provided by \cite{Glavan:2019inb} cannot be realized by a 4-dimensional theory in a consistent manner.

According to the Lovelock theorem, if there indeed exists a novel $4$-dimensional theory with two dofs, the only possibility is that the system cannot be described in a covariant manner. In other words, (a part of) the $4$-dimensional diffeomorphism invariance should be broken.  The best we can do is, therefore, to keep the invariance under the $3$-dimensional spatial diffeomorphism. In this article, we thus explore the EGB gravity in $D=d+1$ dimensions based on the Arnowitt-Deser-Misner (ADM) decomposition. Before taking the $D\rightarrow 4$ ($d\rightarrow 3$) limit, we regularize the Hamiltonian or/and the equations of motion by adding counter terms that respect the $d$-dimensional spatial diffeomorphism invariance but not necessarily the $D$-dimensional spacetime diffeomorphism invariance. We first clarify the reasons why the subtleties arise under the $d\rightarrow 3$ limit. Based on the Hamiltonian formalism, we show that a consistent gravitational theory with two dofs cannot be realized by the $d\rightarrow 3$ limit of the $(d+1)$-dimensional EGB gravity without breaking the temporal diffeomorphism. In order to obtain a consistent $(3+1)$-dimensional theory which is different from general relativity (GR), we need either an additional dof or the violation of the (temporal) diffeomorphism invariance.

In the light of the minimally modified gravity theories (MMGs)~\cite{DeFelice:2015hla,Lin:2017oow,Aoki:2018zcv,Mukohyama:2019unx,DeFelice:2020eju}, we then propose a consistent theory by breaking the temporal diffeomorphism invariance. The theory is defined by a Hamiltonian, purely in $3+1$ dimensions without need for higher dimensions, and possesses the following properties:  (i) the $3$-dimensional spatial diffeomorphism invariance is respected; (ii) the number of local physical dofs in the gravity sector is two; (iii) the theory reduces to GR when $\tilde{\alpha} = 0$; and (iv) each term in the corrections to GR is 4th-order in derivatives. The theory also has the following relation to the EGB gravity: (v) if the Weyl tensor of the spatial metric and the Weyl part of $K_{ik}K_{jl} - K_{il}K_{jk}$, where $K_{ij}$ is the extrinsic curvature, vanish for a solution of the $(d+1)$-dimensional EGB gravity, then the $d\rightarrow 3$ limit of the solution satisfies the equations of motion of the $(3+1)$-dimensional theory. We then argue that the theory is unique up to a choice of  a constraint that stems from a temporal gauge condition, i.e.~$\limitGcal$ which appears in the Hamiltonian, if we assume (i)-(v). Since the GB term in any dimensions is 4th-order in derivatives and most (if not all) of phenomenological consequences of the original suggestion~\cite{Glavan:2019inb} are so far based on solutions in which the Weyl tensor of the $d$-dimensional spatial metric and the Weyl part of $K_{ik}K_{jl} - K_{il}K_{jk}$ vanish, the properties (iv) and (v) make it reasonable to call this theory a theory of $D\rightarrow 4$ EGB gravity. For a convenient choice of $\limitGcal$, we also derive the corresponding Lagrangian. One can use the 4-dimensional theory given by either the Hamiltonian or the Lagrangian to analyze general 4-dimensional spacetimes without assuming any symmetries. 
\\

\section{EGB gravity in $D=d+1$ dimensions}
The $D$-dimensional covariant action of EGB gravity is given by
\begin{align}
S_{\rm EGB}&=\frac{1}{2\kappa^2}\int d^Dx \sqrt{-g}\left[ \mathcal{R}-2\Lambda +\alpha \mathcal{R}_{\rm GB}^2 \right]\,, \\ \nonumber
\mathcal{R}^2_{\rm GB}&=\mathcal{R}^2-4\mathcal{R}^{\mu\nu}\mathcal{R}_{\mu\nu}+\mathcal{R}_{\mu\nu\rho\sigma}\mathcal{R}^{\mu\nu\rho\sigma}\,,
\end{align}
where $g_{\mu\nu}$ is the $D$-dimensional metric, $\mathcal{R}_{\mu\nu\rho\sigma}$ is the associated Riemann curvature tensor, $\kappa$ is the gravitational coupling constant and $\alpha$ is the GB coupling. Since the action has the divergent boundary term under the singular $D\rightarrow 4$ limit after rescaling $\alpha = \tilde{\alpha}/(D-4)$ \cite{Myers:1987yn}, we shall first remove the boundary term by the use of the ADM decomposition.

In the ADM ($D=d+1$) decomposition, it is useful to adopt the Hamiltonian formalism. Following \cite{Teitelboim:1987zz}, the total Hamiltonian of the EGB gravity up to a boundary term is
\begin{align}
\Hd_{\rm tot}=\int d^dx (N\Hdcal_0+N^i \mathcal{H}_i + \lambda^0 \pi_0+\lambda^i \pi_i)\,, \label{Htot}
\end{align}
where $\lambda^0$ and $\lambda^i$ are Lagrange multipliers, and 
\begin{align}
\Hdcal_0&=\frac{\sqrt{\gamma}}{2\kappa^2}
\left[ 2\Lambda - M^{ij}{}_{ij} - \frac{\alpha}{4} 
\delta^{ijkl}_{rstu}M^{rs}{}_{ij}M^{tu}{}_{kl} \right]\,,\nonumber\\
\mathcal{H}_i&= -2\sqrt{\gamma}\gamma_{ik} D_j \Big( \frac{\pi^{jk}}{\sqrt{\gamma}} \Big)\,. 
\end{align}
Here, ($\pi_0$, $\pi_i$, $\pi^{ij}$) are canonical momenta conjugate to ($N$, $N^i$, $\gamma_{ij}$), $D_i$ is the spatial covariant derivative, $\delta^{ijkl}_{rstu}:=4! \delta^{[i}_{\,r} \delta^j_{s} \delta^k_{t} \delta^{l]}_{u}$, $M_{ijkl}:=R_{ijkl}+2K_{i[k}K_{l]j}$, and $K_{ij}$ is understood as the solution of
\begin{eqnarray}\label{exp_pi}
\pi^i_{\,j}=\frac{\sqrt{\gamma}}{2\kappa^2} \Big[ K^i_{\,j}-K\delta^i_j 
- \alpha \delta^{iklr}_{jstu} K^{s}_{\,\,k}
\Big( R^{tu}_{\hspace{.2cm}lr}+\frac{1}{3}K^{t}_{\,\,l}K^{u}_{\,\,r} \Big) \Big]\,. \nonumber
\end{eqnarray}
For $d=3$, the GB contributions identically vanish due to the identity $\delta^{ijkl}_{rstu}\equiv 0$.

The consistent theory that we shall propose, i.e. \eqref{4DEGB} and \eqref{action}, is well-defined for any values of $\tilde{\alpha}$. On the other hand, in order to show the subtleties and ambiguities of the naive $D\to 4$ limit, it suffices and is actually convenient to consider the cases with small $\tilde{\alpha}$ and to expand relevant quantities with respect to $\tilde{\alpha}$. Up to linear order in $\alpha$~\footnote{When (and only when) we adopt expansion with respect to $\alpha$, we shall restrict our consideration to the cases where the GB contributions are of order ${\cal O}(d-3)$. For this reason, by ``linear order in $\alpha$'', we actually mean linear order in $\tilde{\alpha}$.}, we obtain
\begin{widetext}
\begin{align}
\Hdcal_0&=\frac{\sqrt{\gamma}}{2\kappa^2}
\bigg[ 2\Lambda -\Pi-R -\frac{\alpha}{4} \delta^{ijkl}_{rstu}
\Big( R^{rs}{}_{ij} R^{tu}{}_{kl}-2R^{rs}{}_{ij} \Pi^{tu}{}_{kl}
-\frac{1}{3}\Pi^{rs}{}_{ij}\Pi^{tu}{}_{kl} \Big) \bigg]+ {\cal O}(\alpha^2), 
\label{H0_upto1} 
\end{align}
where we have defined
\begin{align}
\Pi^{ijkl}&:=8\kappa^4 \Big( \tilde{\pi}^{i[k}-\frac{1}{d-1}\gamma^{i[k} [\tilde{\pi}]  \Big) 
\Big(\tilde{\pi}^{l]j}-\frac{1}{d-1}\gamma^{l]j } [\tilde{\pi}] \Big),
\quad
\Pi_{ij}:=\Pi^k{}_{ikj}\,, \quad \Pi:=\Pi^i{}_i
\,,
\end{align}
with $\tilde{\pi}^{ij}=\pi^{ij}/\sqrt{\gamma}$ and $[\tilde{\pi}]=\gamma_{ij} \tilde{\pi}^{ij}$. We then replace $\alpha$ with $\tilde{\alpha}/(d-3)$ and split the Hamiltonian into two parts:
\begin{align}
\Hd_{\rm tot} = \Hd_{\rm reg} + \Hd_{\rm Weyl}\,, \quad
\Hd_{\rm reg}=\int d^dx (N\,\Hdcal_{\rm reg}+N^i \mathcal{H}_i + \lambda^0 \pi_0+\lambda^i \pi_i)
\,, \quad
\Hd_{\rm Weyl}=\int d^dx N\, \Hdcal_{\rm Weyl}
\,, \label{H_split}
\end{align}
where
\begin{align}
&\Hdcal_{\rm reg} :=\frac{\sqrt{\gamma}}{2\kappa^2}
\bigg[
2\Lambda-\Pi-R
+\tilde{\alpha} \bigg\{  \frac{4}{d-2} \Big( R_{ij}R^{ij}-2R_{ij}\Pi^{ij}-\frac{1}{3}\Pi_{ij}\Pi^{ij} \Big)
- \frac{d\big( R^2 -2 R\Pi -\frac{1}{3}\Pi^2 \big)}{(d-2)(d-1)} \bigg\} \bigg]
\, + {\cal O}(\tilde{\alpha}^2)
,  \\
&\Hdcal_{\rm Weyl} :=
-\frac{\sqrt{\gamma}}{2\kappa^2} \frac{\tilde{\alpha}}{d-3}
\Big( W_{ijkl}W^{ijkl}-2 W_{ijkl}\Pi^{T\, ijkl}-\frac{1}{3}\Pi^T_{ijkl}\Pi^{T\, ijkl} \Big)
+ {\cal O}(\tilde{\alpha}^2)
, \label{H_Weyl}
\end{align}
and $W_{ijkl}$ and $\Pi^T_{ijkl}$ are irreducible components of the curvature and the tensor $\Pi_{ijkl}$ specified by the traceless conditions $W^k{}_{ikj}=0=\Pi^{T\, k}{}_{ikj}$, namely the Weyl pieces. Because of the relation $\delta^{ijkl}_{rstl}=(d-3)\delta^{ijk}_{rst}$ where $\delta^{ijk}_{rst}:=3!\delta^{[i}_r \delta^j_s \delta^{k]}_t$, the $1/(d-3)$ factors are canceled for the trace pieces and no divergence appears in $\Hd_{\rm reg}$ under the $d\to 3$ limit. We see that only for the Weyl parts, the $d\to3$ limit is not clear since it goes as $0/0$. We shall therefore regularize the Hamiltonian by adding counter terms to cancel these potentially divergent terms before taking the limit. 
\end{widetext}

\section{Number of dofs in $d+1$ dimensions}
For latter convenience, here, we count the number of dofs based on the equations of motion including constraints. The number of dofs is the half of the necessary number of the initial conditions to solve the dynamics of the system, 
\begin{align}
\dot{\gamma}_{ij}&=\delta\Hd_{\rm tot}/\delta\pi^{ij}\,, ~ \dot{\pi}^{ij}=-\delta\Hd_{\rm tot}/\delta\gamma_{ij}\,, \nonumber\\
\dot{N}&=\lambda^0\,, ~ \dot{N}^i=\lambda^i\,, ~ 
\dot{\pi}_0=-\Hdcal_0\,, ~\dot{\pi}_i=-\mathcal{H}_i\,,
\label{dynamical_eq}
\end{align}
with the constraints
\begin{align}
\pi_0&\approx 0\,,~ \pi_i \approx 0\,,~
\mathcal{H}_0\approx 0\,, ~\mathcal{H}_i \approx 0 
\,. \label{constraints}
\end{align}
If we do not take into account the constraints (and gauge conditions discussed below), we generally require $(d+1)(d+2)$ initial conditions to solve \eqref{dynamical_eq}.  However, the system has to satisfy the constraints, which reduce the necessary number of initial conditions. The time derivative of the constraints \eqref{constraints} can be computed by the use of the dynamical equations \eqref{dynamical_eq} and leads to consistency conditions. In the EGB gravity, the consistency conditions of constraints weakly vanish automatically. This implies that the consistency conditions would not lead to additional constraints and that the coefficients $(\lambda^0,\lambda^i,N,N^i)$ would not be determined by the basic equations of the system, meaning the redundancy of description, i.e. the gauge freedom. To fix the redundancy and to solve the dynamics in terms of the given variables (without introducing gauge-invariant variables), we need to impose $2(d+1)$ gauge fixing conditions on variables $(N,N^i,\gamma_{ij},\pi^{ij})$ and consider them as a part of the constraints, and then we would have $4(d+1)$ constraints in total. As a result, the necessary number of the initial conditions turns out to be $(d+1)(d+2)-4(d+1)=(d+1)(d-2)=D(D-3)$, which corresponds to $D(D-3)/2$ dofs. This procedure to count the number of dofs only requires the dynamical equations \eqref{dynamical_eq} and the constraints \eqref{constraints}. Once \eqref{dynamical_eq} and \eqref{constraints} are given, we do not need the concepts of the Hamiltonian and the Poisson bracket.
\\

\section{Subtleties of $D\rightarrow 4$ limit}
As already pointed out by literature there are subtleties in the $D\rightarrow 4~(d\rightarrow 3)$ limit. In particular, it was shown that taking $D\rightarrow 4~(d\rightarrow 3)$, an extra scalar mode shows up  \cite{Lu:2020iav,Kobayashi:2020wqy,Bonifacio:2020vbk,Fernandes:2020nbq,Hennigar:2020lsl} and also the limit is not unique \cite{Lu:2020iav,Kobayashi:2020wqy,Bonifacio:2020vbk}. These facts can be understood by the following observation.

The Weyl decomposition used in \eqref{H_split} is particularly useful to manifest the problematic terms under the $d\rightarrow 3$ limit. Only the ambiguous part of the Hamiltonian under the limit would be the Weyl part $\Hd_{\rm Weyl}$ which generates $0/0$ under the $d\rightarrow 3$ limit. Let us consider a direct product $d$-dimensional space
\begin{align}
\gamma_{ij}dx^i dx^j=\gamma_{ab}dx^a dx^b +r^2(x^a)\delta_{AB}dx^A dx^B
\,, \label{direct_s}
\end{align}
where $\gamma_{ab}$ is a $3$-dimensional spatial metric and we have assumed a flat $(d-3)$-dimensional fiber for simplicity. We then obtain
\begin{align}\label{scaling}
&\gamma^{ij} R_{ij}=\gamma^{ab} R_{ab} + {\cal O}(d-3), \nn
&R^{ij} R_{ij}=R^{ab}  R_{ab}+ {\cal O}(d-3), \nn
&W_{ijkl} W^{ijkl}
=(d-3)\Big( 4R_{ab}R^{ab}-\frac{3}{2}(\gamma^{ab}R_{ab})^2 + \cdots \Big)
\nn
&= {\cal O}(d-3)
\,, 
\end{align}
which leads to
\begin{align}
\lim_{d\rightarrow 3} \int d^d x \sqrt{\gamma} N \bigg( \frac{W_{ijkl}W^{ijkl}}{d-3} \bigg)  = {\rm finite}
\,,
\label{Weyl_limit}
\end{align}
where $\cdots$ in \eqref{scaling} represents terms depending on $\partial_a r$, and similar relations hold for other two terms in \eqref{H_Weyl}. The expressions \eqref{scaling} suggest that, whereas the $(d-3)$-dimensional part does not contribute to $\Hd_{\rm reg}$ in the $d\rightarrow 3$ limit, the dependence on the $(d-3)$-dimensional space survives through $\Hd_{\rm Weyl}$. The $d\rightarrow 3$ limit of \eqref{H_Weyl} depends on the specific form of the $(d-3)$-dimensional metric and thus {\it the $d\rightarrow 3$ limit is not unique}. Note that, as seen in \eqref{H_Weyl}, we should be careful of not only the Weyl tensor but also the other Weyl piece $\Pi^T_{ijkl}$. There is ambiguity of the $d\rightarrow 3$ limit coming from the $(d-3)$-dimensional part of the canonical momentum as well.

We then return to generic $(d+1)$-dimensional spacetime. We need to determine the procedure of the limit to ``define'' the $D\rightarrow 4$ EGB theory because the limit is not unique. 
Except the Weyl part $\Hd_{\rm Weyl}$, we may naturally define the $d\rightarrow 3$ limit by identifying the $d$-dimensional tensors with the $3$-dimensional ones. 
As stated, the $D\rightarrow 4$ theory depends on how to regularize the Weyl terms. Since the finite contribution of \eqref{Weyl_limit} arises from the $(d-3)$-dimensional part, the $d\rightarrow 3$ limit with non-vanishing Weyl terms generically implies existence of an additional dof; for instance, time derivatives of $r(x)$ survives after taking the limit from the direct product space \eqref{direct_s}, and the resultant $D\rightarrow 4$ EGB gravity has a scalar dof as shown by \cite{Lu:2020iav,Kobayashi:2020wqy,Bonifacio:2020vbk}. Apart from the non-uniqueness of this scalar-tensor theory, it is not a new theory but an ill-defined strongly coupled subset of the Horndeski theory \cite{Kobayashi:2020wqy}. Moreover, looking at the gravitational scattering amplitudes and by very general arguments, it is shown that there is no new $D\to4$ EGB scalar-tensor theory \cite{Bonifacio:2020vbk}.

Therefore, there are infinite number of ways to have finite contribution in the right hand side of (\ref{Weyl_limit}) which all imply increase in the number of dofs. In the simplest case with one scalar dof, there is not any new theory. However, still there is another possibility: removing these problematic Weyl terms by adding appropriate counter terms. In this regard, we prevent the appearance of extra dofs while the resultant theory would be Lorentz-violating in general since the counter terms are only invariant under spatial diffeomorphism. In a particular case and motivated by the specific ansatz \eqref{scaling}, a possible limit of the Weyl term \eqref{Weyl_limit} without information of $(d-3)$-dimensional part (without extra dofs) would be $\int d^3 x \sqrt{\gamma} N (4R_{ij}R^{ij}-3R^2/2)$. However, if we further assume the $4$-dimensional diffeomorphism invariance then this limit concludes that $\Hd_{\rm Weyl}$ cancels the GB contribution in $\Hd_{\rm reg}$. The resultant theory is nothing but GR.

Therefore, a natural possibility to have a novel theory with two dofs is the following $d\rightarrow 3$ limit: {\it we first remove all Weyl pieces by adding counter terms that are invariant under spatial diffeomorphism and then take the $d\rightarrow 3$ limit by identifying all $d$-dimensional tensors with $3$-dimensional ones}. This way, we can practically take the $d\rightarrow 3$ limit not only for a functional but also for tensors\footnote{In practice, we can set the Weyl pieces to vanish before taking the limit. In principle, we achieve this goal by adding counter terms to cancel the subtle Weyl terms $\Hdcal_{\rm Weyl}$.}.

However, there still exists an ambiguity to define the $d\rightarrow 3$ theory as we have two options: 1) we take the $d\rightarrow 3$ limit of the Hamiltonian and then derive the $(3+1)$-dimensional Hamilton equations, or 2) we first derive the Hamiltonian equations in $d+1$ dimensions and then take the $d\rightarrow 3$ limit. The resultant theories do not coincide because the limit and the (functional) derivative do not commute, in general. To see this fact explicitly, notice that before taking the $d\rightarrow 3$ limit we add counter terms to remove the Weyl pieces such as 
\begin{align}
\int d^d x \sqrt{\gamma} \, N \bigg( \frac{ W_{ijkl}W^{ijkl} }{d-3} \bigg)\,,
\label{Weyl_limit1}
\end{align}
but the variation of it (before the regularization and the limit) is
\begin{align}
& \delta \int d^d x \sqrt{\gamma} N \bigg( \frac{ W_{ijkl}W^{ijkl} }{d-3} \bigg) \nonumber \\ \nonumber
& = 4 \int d^dx \sqrt{\gamma} \big( 2 D_k N + N D_k \big) \bigg( \frac{D_lW^{kjil} }{d-3} \bigg) \delta\gamma_{ij} \\ \nonumber
& + \mbox{terms including Weyl tensor itself} \,,
\end{align}
where we have only shown the terms that include spatial derivative of the Weyl tensor. Using $(d-2)D_lW^{kjil} = - (d-3) C^{ijk}$, where $C_{ijk}$ is the Cotton tensor, and taking the $d\rightarrow 3$ limit, we find a finite contribution
\begin{align}
&\lim_{d\rightarrow 3} \left[ \delta \int d^d x \sqrt{\gamma} N \bigg( \frac{W_{ijkl}W^{ijkl}}{d-3} \bigg) + \mbox{counter terms} \right] 
\nn
&= -4\int d^3 x \sqrt{\gamma} \big( 2 D_k N + N D_k \big) C^{ijk} \delta \gamma_{ij} \,. 
\label{Weyl_limit2}
\end{align}
We thus discuss these two possibilities in order.
\\

\section{No covariant $D\rightarrow 4$ EGB with two dofs}
Let us now study the first possibility, namely the $d\rightarrow 3$ limit at the level of the Hamiltonian. The $3$-dimensional Hamiltonian is given by
\begin{align}
\limitH_{\rm tot}=\int d^3x (N \, \limitHcal_0+N^i \mathcal{H}_i + \lambda^0 \pi_0+\lambda^i \pi_i) \,,
\end{align}
where $\limitHcal_0=\lim_{d\rightarrow 3} \Hdcal_{\rm reg}$ and $\mathcal{H}_i$ takes the standard form. Since we know the explicit form of the Hamiltonian, we can straightforwardly count the number of dofs. We find $\{ \limitHcal_0(x), \limitHcal_0(y) \}$ does not weakly vanish at ${\cal O}(\tilde{\alpha})$, which shows that the temporal diffeomorphism invariance is broken. Other constraints $\mathcal{H}_i \approx 0, \pi_0\approx0, \pi_i \approx 0$ clearly commute with $\limitHcal_0$. Hence, the situation does not change even if we consider a linear combination of the constraints.

We then consider the $d\rightarrow 3$ limit of the Hamilton equations. The $(3+1)$-dimensional dynamical equations are
\begin{align}
\dot{\gamma}_{ij}&={}^3 \! F_{ij} \,, ~ \dot{\pi}^{ij}={}^3 \! G^{ij}= {}^3 \! \bar{G}^{ij} +{}^3 \! \delta G^{ij} \,,~
\nn
\dot{N}&=\lambda^0\,, ~ \dot{N}^i=\lambda^i 
\,, ~
\dot{\pi}_0=-\limitHcal_0 \,, ~\dot{\pi}_i=-\mathcal{H}_i \,,
\label{dynamical_eq2}
\end{align}
where ${}^3 \!  F_{ij} = \delta\limitH_{\rm tot}/\delta\pi^{ij}$, ${}^3 \! \bar{G}^{ij} = -\delta\limitH_{\rm tot}/\delta\gamma_{ij}$ and
\begin{align}
&{}^3 \! \delta G^{ij} =
\frac{\tilde{\alpha}}{2\kappa^2} \Biggl[ 4 \big( 2 D_k N + N D_k \big) C^{ijk} 
\\
&+ \frac{16\kappa^8}{3}N\biggl\{ -\frac{4}{3}\tilde{\pi}^{ij} \left( 5[\tilde{\pi}^3]-3 [\tilde{\pi}][\tilde{\pi}^2] \right)
+12\tilde{\pi}^4{}^{ij}-8\tilde{\pi}^3{}^{ij}[\tilde{\pi}]
\nn
& 
 + \frac{2}{3}\gamma^{ij}\left(-[\tilde{\pi}]^4-3[\tilde{\pi}^2]^2+3 [\tilde{\pi}]^2[\tilde{\pi}^2]+[\tilde{\pi}][\tilde{\pi}^3] \right) \biggl\} \Biggl]
+ {\cal O}(\tilde{\alpha}^2) , \nonumber
\end{align}
with $\tilde{\pi}^n{}^{ij}=\tilde{\pi}^i_{i_1} \tilde{\pi}^{i_1}_{i_2} \cdots \tilde{\pi}^{i_n j}$ and $[\tilde{\pi}^n]=\tilde{\pi}^{nij}\gamma_{ij}$\footnote{We have vanished the terms proportional to $W_{ijkl},\Pi^T_{ijkl}$ and $D^i\Pi^T_{ijkl}$ with the $\tilde{\alpha}/(d-3)$ coefficients before the $\lim_{d\rightarrow 3}$ limit in order to obtain ${}^3 \! \delta G^{ij}$. }.
Due to the term ${}^3 \! \delta G^{ij}$, this system is clearly inequivalent to the former case. The constraints are
\begin{align}
\pi_0&\approx 0\,,~ \pi_i \approx 0\,,~
\limitHcal_0\approx 0\,, ~\mathcal{H}_i \approx 0 
\,.
\end{align}
The time derivative of the Hamiltonian constraint $\limitHcal_0\approx 0$ is now computed by the chain rule, 
\begin{align}
\dot{\limitHcal}_0(x) 
&=\int d^3 z \left[ \frac{\delta \limitHcal_0(x)}{\delta \gamma_{ij}(z)} {}^3 \! F_{ij}(z) 
+\frac{\delta \limitHcal_0(x)}{\delta \pi^{ij}(z)} {}^3 \! G^{ij}(z)  \right]\,.\nonumber
\end{align}
We then find that $\dot{\limitHcal}_0$ does not weakly vanish at ${\cal O}(\tilde{\alpha})$, meaning that the temporal diffeomorphism invariance is broken again. 

We also comment on another subtlety in the second approach. If there exists a Hamilton functional that reproduces the equations \eqref{dynamical_eq2}, the functions ${}^3\! F_{ij}$ and ${}^3 \! G^{ij}$ must satisfy the integrability conditions $\delta {}^3 \! F_{ij}(x) / \delta \pi^{kl}(y) = \delta {}^3 \! F_{kl}(y) / \delta \pi^{ij}(x)$, $\delta {}^3 \! F_{ij}(x) / \delta \gamma_{kl}(y) = \delta {}^3 \! G^{kl}(y) / \delta \pi^{ij}(x)$, $\delta {}^3 \! G^{ij}(x) / \delta \gamma_{kl}(y) = \delta {}^3 \! G^{kl}(y) / \delta \gamma_{ij}(x)$. These conditions do not hold for ${}^3\! F_{ij}$ and ${}^3 \! G^{ij}$, meaning that the set of equations \eqref{dynamical_eq2} does not define a Hamiltonian flow for the given set of variables. 

In summary, we cannot obtain a covariant $D\rightarrow 4$ EGB gravity with two dofs when we remove the potentially divergent Weyl pieces by adding counter terms before taking the $D\to 4$ ($d\to 3$) limit. Although one may define the right-hand-side of \eqref{Weyl_limit} by a finite quantity, it implies that this finite quantity has information about the $(d-3)$-dimensional space and then the resultant theory must have an additional dof since $\cdots$ in \eqref{scaling} and corresponding parts for the other two terms in \eqref{H_Weyl} contain the kinetic term of $r$. We therefore need to introduce either an additional dof or violation of the (temporal) diffeomorphism invariance to obtain a novel $4$-dimensional EGB gravity. This conclusion is consistent with the Lovelock theorem. 
\\

\section{A consistent $D\rightarrow 4$ EGB with two dofs}
From now on, we shall construct a consistent theory in ($3+1$)-dimensions without relying on the expansion with respect to $\tilde{\alpha}$ (or $\alpha$). When one renounces the temporal diffeomorphism invariance, one can obtain gravitational theories with two dofs, dubbed MMGs \cite{DeFelice:2015hla,Lin:2017oow,Aoki:2018zcv,Mukohyama:2019unx,DeFelice:2020eju} (see also the cuscuton theories \cite{Afshordi:2006ad,Iyonaga:2018vnu}). A consistent $D\rightarrow 4$ EGB gravity with two dofs can be formulated in the framework of MMGs.

Since the equations of motion of the $D\rightarrow 4$ EGB gravity have not been obtained, the ``solutions'' of the $D\rightarrow 4$ EGB gravity so far were found by the $D\rightarrow 4$ limit of the solutions of the $D$-dimensional EGB gravity~\cite{Glavan:2019inb}. Due to the $D$-dimensional diffeomorphism invariance, at least locally, any solutions of \eqref{Htot} are solutions of the gauge-fixed Hamiltonian, 
\begin{align}
\Hd'_{\rm tot}= \Hd_{\rm tot}+\int d^dx \lambda_{\rm GF} \Gdcal(\gamma_{ij},\pi^{ij})
\,.
\end{align}
Here, the gauge-fixing term is defined by the requirement that $\{\Hdcal(x),\Gdcal(y) \}$ does not have a non-trivial kernel as well as $\{\mathcal{H}_i(x),\Gdcal(y) \} \approx 0$, $\{\pi_0(x),\Gdcal(y) \} \approx 0$ and $\{\pi_i(x),\Gdcal(y) \} \approx 0$, i.e.~we have only fixed the temporal gauge. To remove the problematic Weyl terms, we add the counter term $\Hd_{\rm ct}:=-\Hd_{\rm Weyl}$,
\begin{align}
\Hd_{\rm tot}'' &= \Hd_{\rm tot}+ \int d^dx \lambda_{\rm GF} \Gdcal(\gamma_{ij},\pi^{ij}) + \Hd_{\rm ct} 
\nn 
&=\Hd_{\rm reg} + \int d^dx \lambda_{\rm GF} \Gdcal(\gamma_{ij},\pi^{ij})
\end{align}
and then take the $d\rightarrow 3$ limit of the gauge-fixed Hamiltonian\footnote{One may take the $d\rightarrow 3$ limit at the level of gauge-fixed equations of motion; however, the $d\rightarrow 3$ limit of the equations would not satisfy the integrability conditions, similarly to the case without the gauge-fixing.},  keeping $\lim_{d\to 3}\Gdcal$ as a constraint.

The theory constructed in this way is defined purely in $3+1$ dimensions by the Hamiltonian,
\begin{eqnarray}
H^{\rm 4D}_{\rm EGB}&=& \int d^3 x 
(N \limitHcal_0+N^i \mathcal{H}_i + \lambda^0 \pi_0+\lambda^i \pi_i+\lambda_{\rm GF} \limitGcal)\,, \nonumber\\
\limitHcal_0&=& \frac{\sqrt{\gamma}}{2\kappa^2} \Big[ 2\Lambda - \mathcal{M} + \tilde{\alpha} 
\Big(4 \mathcal{M}_{ij} \mathcal{M}^{ij}-\frac{3}{2} \mathcal{M}^2 \Big) \Big]\,,\nonumber\\
\mathcal{H}_i&=& -2\sqrt{\gamma}\gamma_{ik} D_j \Big( \frac{\pi^{jk}}{\sqrt{\gamma}} \Big)\,,
 \label{4DEGB}
\end{eqnarray}
where $\mathcal{M}_{ij}:=R_{ij}+\mathcal{K}^k_{\, k} \mathcal{K}_{ij}-\mathcal{K}_{ik}\mathcal{K}^k_{\, j}$, $\mathcal{M}:=\mathcal{M}^i_{\, i}$, $\mathcal{K}_{ij}$ is understood as the solution of
\begin{align} \label{def_K}
\pi^i_{\,j}= \frac{\sqrt{\gamma}}{2\kappa^2} &\Big[ \mathcal{K}^i_{j}-\mathcal{K}\delta^i_{j} 
-\frac{8}{3}\tilde{\alpha}
\delta^{ikl}_{jrs}\mathcal{K}^r_{k}
\\ \nonumber 
&\times \Big( R^s_{\,l}-\frac{1}{4}\delta^s_{\,l} R + \frac{1}{2} \big( \mathcal{M}^s_{\,l} 
- \frac{1}{4} \delta^s_{\,l} \mathcal{M} \big) \Big) \Big],
\end{align}
with $\delta^{ijk}_{rst}:=3!\delta^{[i}_{\,r}\delta^{j}_{\,s}\delta^{k]}_{\,t}$, and the constraint $\limitGcal$ is required to satisfy the condition that $\{\limitHcal_0(x),\limitGcal(y) \}$ does not have a non-trivial kernel as well as $\{\mathcal{H}_i(x),\limitGcal(y) \} \approx 0$, $\{\pi_0(x),\limitGcal(y) \} \approx 0$ and $\{\pi_i(x),\limitGcal(y) \} \approx 0$. 

It is easy to see that the system described by the Hamiltonian \eqref{4DEGB} has the properties (i)-(v) listed in the introduction and summary section. Since the momentum constraints are first-class and satisfy the standard algebra, (i) holds. Since there are enough number of the constraints, i.e. 6 first-class constraints
\begin{align}
\pi_i \approx 0\,, \quad \mathcal{H}_i \approx 0 \,,
\end{align}
and $2\times 2$ second-class constraints
\begin{align}
\pi_0 \approx 0\,, \quad \limitHcal_0 \approx 0 \,, \quad
\limitGcal \approx 0 \,, \quad \dot{\limitGcal} \approx 0
\,,
\end{align}
(ii) also holds. From the form of the Hamiltonian \eqref{4DEGB}, (iii) and (iv) are obvious. Finally, the construction of the Hamiltonian ensures that (v) also holds. The Hamiltonian \eqref{4DEGB} can thus be interpreted as a consistent $D\rightarrow 4$ EGB gravity with two dofs. 

If we demand (i)-(iii) but not (iv)-(v) then one can add arbitrary $N$- and $N^i$-independent, spatial scalar density to the $\tilde{\alpha}$ part of $\limitHcal_0$, as far as the constraint $\limitGcal$ still satisfies the above mentioned conditions. Since the conformal flatness is characterized by vanishing Weyl tensor in $d>3$ dimensions and by vanishing Cotton tensor in $3$ dimensions, (v) then restricts possible additional terms in the $\tilde{\alpha}$ part of $\limitHcal_0$ to polynomials of the Cotton tensor and its covariant spatial derivatives. Since such polynomials are six order or higher in derivatives, they are excluded by (iv). In summary, if we demand (i)-(v) then the only possible Hamiltonian is \eqref{4DEGB} up to the choice of $\limitGcal$. 

In this framework, the constraint $\limitGcal\approx 0$ is a part of the definition of the theory. Hence, the $D\rightarrow 4$ EGB gravity satisfying (i)-(v) is unique only up to the choice of $\limitGcal(\gamma_{ij},\pi^{ij})$. The theory does not enjoy the full diffeomorphism invariance but is invariant under spatial diffeomorphism.

A useful choice compatible with cosmology and static configurations is $\limitGcal=\sqrt{\gamma} D^2 [\tilde{\pi}]$~\cite{Aoki:2018brq}. This gauge condition reduces to the constant mean curvature slice $K=K(t)$ when we take the GR limit $\tilde{\alpha} \to 0$. Adopting this choice and performing the Legendre transformation, we obtain the Lagrangian density that corresponds to \eqref{4DEGB},
\begin{align}
\mathcal{L}^{\rm 4D}_{\rm EGB}&=\frac{1}{2\kappa^2} ( -2\Lambda +\mathcal{K}_{ij}\mathcal{K}^{ij}-\mathcal{K}^i_{\,i}\mathcal{K}^j_{\,j}+R+\tilde{\alpha} R^2_{\rm 4DGB} )\,, \nonumber\\
R^2_{\rm 4DGB}&=-\frac{4}{3}\left(8R_{ij}R^{ij}-4R_{ij}\mathcal{M}^{ij}-\mathcal{M}_{ij}\mathcal{M}^{ij}\right) \nonumber\\
&+\frac{1}{2}\left( 8R^2 -4 R\mathcal{M} -\mathcal{M}^2 \right)\,, 
\label{action}
\end{align} 
where $\mathcal{K}_{ij}$ is given by $\mathcal{K}_{ij}=K_{ij}-\frac{1}{2N}\gamma_{ij}D^2 \lambda_{\rm GF}=\frac{1}{2N}( \dot{\gamma}_{ij}-2D_{(i}N_{j)}-\gamma_{ij}D^2 \lambda_{\rm GF} )$.\footnote{Even when $R^2_{\rm 4DGB}$ is replaced with a spatial scalar function $f(\gamma_{ij},R_{ij},\mathcal{K}_{ij},D_i)$, the theory still satisfies the conditions (i), (ii) and (iii), in general. The conditions (iv) and (v) determines the form of $f$.} 
The condition (v) ensures that the $d\rightarrow 3$ limit of the solutions of the $(d+1)$-dimensional EGB gravity with a conformally flat spatial metric and a vanishing Weyl part of $K_{ik}K_{jl} - K_{il}K_{jk}$ are also solutions of the $(3+1)$-dimensional theory, either \eqref{4DEGB} or \eqref{action}, provided that a gauge condition $\Gdcal\approx 0$ satisfying $\lim_{d\to 3}\Gdcal = \limitGcal$ is imposed while taking the limit. 

In particular the FLRW and black hole solutions that were found in \cite{Glavan:2019inb} are solutions of the $(3+1)$-dimensional theory that we have defined here as can be checked explicitly (see \cite{Aoki:2020iwm} for the case of FLRW solution and other cosmological implications of the consistent theory). On the other hand, the $d\rightarrow 3$ limits of other types of $(d+1)$-dimensional solutions are not guaranteed to be solutions of the $(3+1)$-dimensional theory. For instance, the spatial metric for the gravitational waves is not conformally flat and, because of our regularization scheme, the spatial higher derivatives would show up in the equation of motion of gravitational waves so that the corresponding dispersion relation is modified to the form $\omega^2=c_T^2k^2+\beta k^4/M_*^2$ \cite{Aoki:2020iwm}. This a direct evidence that the naive $D\to 4$ limit leads to an inconsistent result and the Lorentz violation is inevitable from the consistency of the theory.

The Lorentz violation is in the gravity sector and suppressed by $\tilde{\alpha}$. At the classical level, we can assume that the matter action respects the local Lorentz invariance. At the quantum level, the Lorentz violation in the gravity sector should percolate to the matter sector via graviton loops. However, as far as the matter (i.e. the standard model) is minimally coupled to the metric, such a Lorentz violation in the matter sector is suppressed not only by $\tilde{\alpha}$ but also by negative powers of $M_{\rm Pl}^2$. In this sense Lorentz violation is under control. It is certainly interesting to investigate phenomenological implications of the Lorentz violation in the matter sector induced by graviton loops.

It is also worth mentioning some similarities and differences between the theory \eqref{action} and Ho\v{r}ava-Lifshitz gravity~\cite{Horava:2009uw}. Both theories break the temporal diffeomorphism invariance and exhibit nonlinear dispersion relations. In both theories, the action contains only up to two time derivatives but includes higher spatial derivatives, and this distinction between time and space uniquely defines the the preferred frame. However, the number of dofs is different: Ho\v{r}ava-Lifshitz gravity without extra structures (such as extra $U(1)$ symmetry~\cite{Horava:2010zj,daSilva:2010bm}) has a scalar dof in addition to the standard tensor dofs as a result of the violation of the temporal diffeomorphism invariance whereas the theory \eqref{action} only has the tensor dofs. The extra dof(s) in the theory \eqref{action} is eliminated by the special structure of $\mathcal{K}_{ij}$ but there is no a priori reason why this structure could be stable against quantum corrections. If quantum corrections change the structure of $\mathcal{K}_{ij}$ then 
extra dof(s) may emerge. This is not a problem from the viewpoint of the low energy effective field theory since the extra dof(s) should be heavy and can be integrated out at low energies. However, this probably means that the theory \eqref{action} and its extensions~\footnote{Relaxing condition (iv) listed in the introduction and summary section, we can add the Cotton tensor square term which provides a $k^6$ term in the dispersion relation that is the characteristic feature of the Ho\v{r}ava-Lifshitz gravity.} would be non-renormalizable. On the other hand, the projectable version of Ho\v{r}ava-Lifshitz gravity was recently proved to be renormalizable~\cite{Barvinsky:2015kil}.

Finally, we recall that the scalar-tensor description of the $D\rightarrow 4$ EGB gravity in the literature suffers from infinite strong coupling around the FLRW spacetime \cite{Kobayashi:2020wqy} and that the FLRW spacetime cannot be trusted as a solution of the scalar-tensor description.
On the other hand, our theory of the $D\rightarrow 4$ EGB gravity is free from such a pathology and can consistently describe physics around any solutions such as the FLRW background since the number of dofs is two at nonlinear orders. Therefore, the FLRW and black hole solutions found in \cite{Glavan:2019inb} should be counted as solutions of the consistent theory defined by \eqref{4DEGB} or \eqref{action} and not as the solutions of the equations of motion that are naively (and inconsistently) suggested in \cite{Glavan:2019inb}. The equations of motion that are obtained in \cite{Glavan:2019inb} are divergent in general which is explicitly confirmed in \cite{Arrechea:2020evj} in the case of second order perturbations. We leave the analysis of rotating black holes and quasi-normal modes to future works. \\

{\bf Acknowledgments.}
K.A. and M.A.G. acknowledge the xTras package~\cite{Nutma:2013zea} which was used for tensorial calculations. 
The work of K.A. was supported in part by Grants-in-Aid from the Scientific Research Fund of the Japan Society for the Promotion of Science, No.~19J00895 and No.~20K14468. 
The work of M.A.G. was supported by Japan Society for the Promotion of Science Grants-in-Aid for international research fellow No. 19F19313.
The work of S.M. was supported by Japan Society for the Promotion of Science Grants-in-Aid for Scientific Research No.~17H02890, No.~17H06359, and by World Premier International Research Center Initiative, MEXT, Japan.

\end{document}